\documentclass[letter]{aa}

\usepackage{graphicx}
\usepackage{mathtools}
\usepackage{amsmath}
\usepackage{fancyref}
\usepackage{longtable}
\usepackage{rotating}
\usepackage{pdflscape}
\usepackage{enumerate}
\usepackage{color}
\usepackage[citecolor=blue, urlcolor=blue, linkcolor=blue, colorlinks=true]{hyperref}
\usepackage{epsfig}
\usepackage{epstopdf}
\usepackage{natbib}
\usepackage{gensymb}
\usepackage{txfonts}
\newcommand{\lo}    {$L_{\odot}$}

\newcommand{\lint} {$L_{\rm int}$}
\newcommand{\mo}    {$M_{\odot}$}

\newcommand{\mj}    {$M_\mathrm{Jup}$}
\newcommand{\ceo}   {C$^{18}$O\,}

\begin{document}

   \title{IRAM\,04191+1522: A compact proto-brown dwarf binary candidate\thanks{Based on observations obtained at the ALMA Observatory under programs 2017.1.00551.S and 2016.1.00039.S.}}
   \author{N. Hu\'elamo\inst{1}
          \and 
          I. de Gregorio-Monsalvo\inst{2}
          \and
          Aina Palau\inst{3}
          \and
          C. Carrasco-Gonz\'alez\inst{3}    
          \and 
          A. Ribas\inst{4}
          \and 
          H. Bouy\inst{5}
         \and
         R. Pandey\inst{3}
         \and
        D. Barrado\inst{1}
         \and
        N. Otten\inst{6}
        \and
        V.D. Ivanov\inst{7}
        \and
        M.F. Sterzik\inst{7}
       \and
       M. Dunham\inst{8}
       \and
       L. A. Zapata\inst{3}
       \and
       E. Pantin\inst{9}
       \and 
       E. Macias\inst{7}
          }
    \institute{Centro de Astrobiolog\'{\i}a (CAB), CSIC-INTA,  ESAC Campus, Camino bajo del Castillo s/n, E-28692 Villanueva de la Ca\~nada, Madrid, Spain\\
    \email{nhuelamo@cab.inta-csic.es}
    \and
    European Southern Observatory, Alonso de Cordova 3107, Casilla 19, Vitacura, Santiago, Chile
    \and
    Instituto de Radioastronom\'ia y Astrof\'isica, Universidad Nacional Aut\'onoma de M\'exico, Antigua Carretera a P\'atzcuaro 8701, Ex-Hda. San Jos\'e de la Huerta, 58089 Morelia, Michoac\'an, M\'exico   
    \and  
    Institute of Astronomy, University of Cambridge, Madingley Road, CB3 0HA Cambridge, UK
    \and
    Laboratoire d'Astrophysique de Bordeaux, Univ. Bordeaux, CNRS, B18N, all\'ee Geoffroy Saint-Hilaire, 33615 Pessac, France
    \and
    Department of Physics, Maynooth University, Maynooth, Co.Kildare, Ireland
    \and
    European Southern Observatory, Karl-Schwarzschild-Strasse 2, 85748 Garching bei M\"unchen, Germany
    \and
    Department of Physics, Middlebury College, Middlebury, VT 05753, USA
    \and
    Universit\'e Paris-Saclay, Universit\'e Paris Cit\'e, CEA, CNRS, AIM, 91191, Gif-sur-Yvette, France   
    }   
           \date{Received ; accepted }

 \abstract
   {Very low luminosity objects (VeLLOs) in nearby star-forming regions have been identified as promising proto-brown dwarf candidates, and
   studying their multiplicity can shed light on the dominant formation mechanism of substellar objects.
   }
    {Our aim is to study the multiplicity of the VeLLO IRAM\,04191+1522. 
  }
    {We obtained 0.89\,mm ALMA observations with a very extended configuration that achieved an angular resolution of $\sim$0$\farcs$04 (6\,au at 140\,pc). We complemented these data with new VLA observations and ALMA archival data at 1.3\,mm.
   }
   {We resolved IRAM04191+1522 into a close binary candidate for the first time. The binary is detected in the ALMA continuum data with a projected separation of $\sim$80\,mas (or 11\,au at a distance of 140\,pc). 
   The two sources are oriented in the east-west direction, with the eastern component being brighter and more extended than the western one, which is marginally resolved. 
   Analysis of \ceo(2-1) archival data revealed gaseous material in rotation around the binary, presumably from a circumbinary disk with $\sim$27\,au of radius centered on the faintest ALMA component. A fit of the position-velocity diagram allowed us to estimate a total dynamical mass for the system of 50$\pm$40\,\mj. Therefore, we classify IRAM04191 as a tight proto-brown dwarf binary candidate. The VLA data revealed the detection of a single object closer to the western ALMA source with a spectral index consistent with a radio jet.}
  {}
  \keywords{Brown Dwarf Formation  -- binaries -- individual: IRAM\,04191+1522}
\maketitle
\nolinenumbers

\section{Introduction}

Very low luminosity objects (VeLLOs), discovered by the {\em Spitzer} space telescope, are defined as embedded objects with internal luminosities (\lint) below 0.1\,\lo\,\citep{diFrancesco2007}. They have been identified as promising proto-brown dwarf (proto-BD) candidates, that is, BDs at the earliest stages of their evolution when they are still embedded in their parental clouds \citep[see review by][]{Palau2024_review}. 
The multiplicity of proto-BDs is an important outcome of substellar formation theories, and its study can help shed light on their dominant formation mechanism \citep[e.g.][]{Bate2012,Offner2016}. 

We designed a pilot project to study the multiplicity of VeLLOs with ALMA that obtained observations for IRAM\,04191+1522 (IRAM04191, hereafter). IRAM04191 is a Class~0 object located in the southern part of the Taurus molecular cloud displaying 
a collimated CO(2-1) bipolar outflow 
and an ionized jet \citep{Andre1999}. It was detected by the {\em Spitzer} telescope at wavelengths shorter than 60\,$\mu$m and classified as a VeLLO based on its \lint\ of 0.08$\pm$0.04\lo\, \citep{Dunham2006}. 
In this letter, we present unprecedented angular resolution ALMA observations of IRAM04191 at 0.89\,mm that have allowed us to resolve the object into a tight binary candidate. 
We complement these data with new Very Large Array (VLA) observations and ALMA Band~6 (1.3\,mm) archival data. 

\section{Observations and data reduction}\label{data}

\subsection{Band\,7 ALMA observations}\label{alma:b7}

As part of the ALMA Cycle~5 program 2017.1.00551.S (PI Hu\'elamo), IRAM04191 was observed with the ALMA 12m-array on November 24, 2017. The data were obtained with 49 antennas in single field dual polarization mode. Band~7 (334.9\,GHz, 0.89\,mm) was used, with two basebands centered at 335.9 and 334.0 GHz, respectively, both with a width of 1.875 GHz and a channel spacing of 31 MHz. 
We also placed two spectral windows respectively centered at the frequency that contains the CO(3-2) and H$^{13}$CO+(4-3) transitions to obtain serendipitous detections of gas emission associated with outflow and disk phenomena. The observations were performed with an average precipitable water-vapor column of $\sim$0.6\,mm. The antenna baselines ranged from 92\,m to 8.5\,km, providing a maximum recoverable scale of 0\farcs6. 
The objects QSO\,J0237+2848 and QSO\,J0440+1437 were respectively used as a bandpass calibrator and a phase calibrator. The former
was used to scale the flux density. 
The science time on-source was 16 minutes.  
We used the automatic self-calibration pipeline \texttt{auto\_selfcal}\footnote{\url{https://github.com/jjtobin/auto_selfcal}}
 to self-calibrate the visibilities and perform continuum imaging \citep{auto_selfcal}, changing the \texttt{minsnr\_to\_proceed} parameter to be 1.5 instead of 2.95. 
 The pipeline was executed using the Common Astronomy Software Applications package \citep[CASA;][]{McMullin2007_casa}
 version 6.6.4. The data self-calibrated successfully, with the phase-only self-calibration converging down to the scan-length solution interval. 
 We analyzed data with different imaging options (see Appendix \ref{appendix_b7data}), selecting the image with a robust $-0.25$ and a cut in $uv$ range of $>$ 200 $k\lambda$ as our fiducial image showing a synthesized beam size of 33$\times$25\,mas and position angle (PA) of 42\degree. 

\subsection{VLA observations}

We observed IRAM04191 with the VLA of the National Radio Astronomy Observatory (NRAO)
in its A configuration at the K, Ka, and Q bands (project 24B-250). All observations were performed during the second semester of 2024: October 22 for the Q band and December~4 and 5 for Ka and K, respectively. At all bands, we used the standard continuum observing setup that allows one to use a total of 8\,GHz in bandwidth for each band. 
For all bands, 3C147 was used as a flux calibrator,  while 3C84 was used as the bandpass calibrator, and J0409+1217 was used as a complex gain calibrator. At each band, we used cycles between the target and complex gain calibrator as recommended for the A configuration by the NRAO. Data were calibrated by the NRAO staff using CASA v6.6.1 and the VLA pipeline. We tried self-calibration of the data, but it did not work due to the faintness of the source. Images were made using \texttt{tclean} from the calibrated data and exploring different visibility weightings (natural, uniform, and briggs). 
We produced a single image combining all three bands (KKaQ) and using multi-frequency synthesis. We worked with the image with a robust parameter of one, showing a synthesized beam size of 0\farcs08$\times$0\farcs07 and a PA of -33\degree. 

\subsection{Band\,6 ALMA archival observations}

We complemented our 0.89\,mm data with public ALMA archival data obtained in Band~6 (220.6\,GHz, 1.3\,mm) under projects 2016.1.00039.S (P.I. Dunham) and 2016.1.01284.S (P.I. Maury). We note that the former was observed with an angular resolution comparable to our 0.89\,mm data. The description of these observations and the data reduction are included in Appendix~\ref{App:ALMA_archival_data}.

\section{Main results}\label{results}

\begin{figure*}[t!]
\includegraphics[width=18cm]{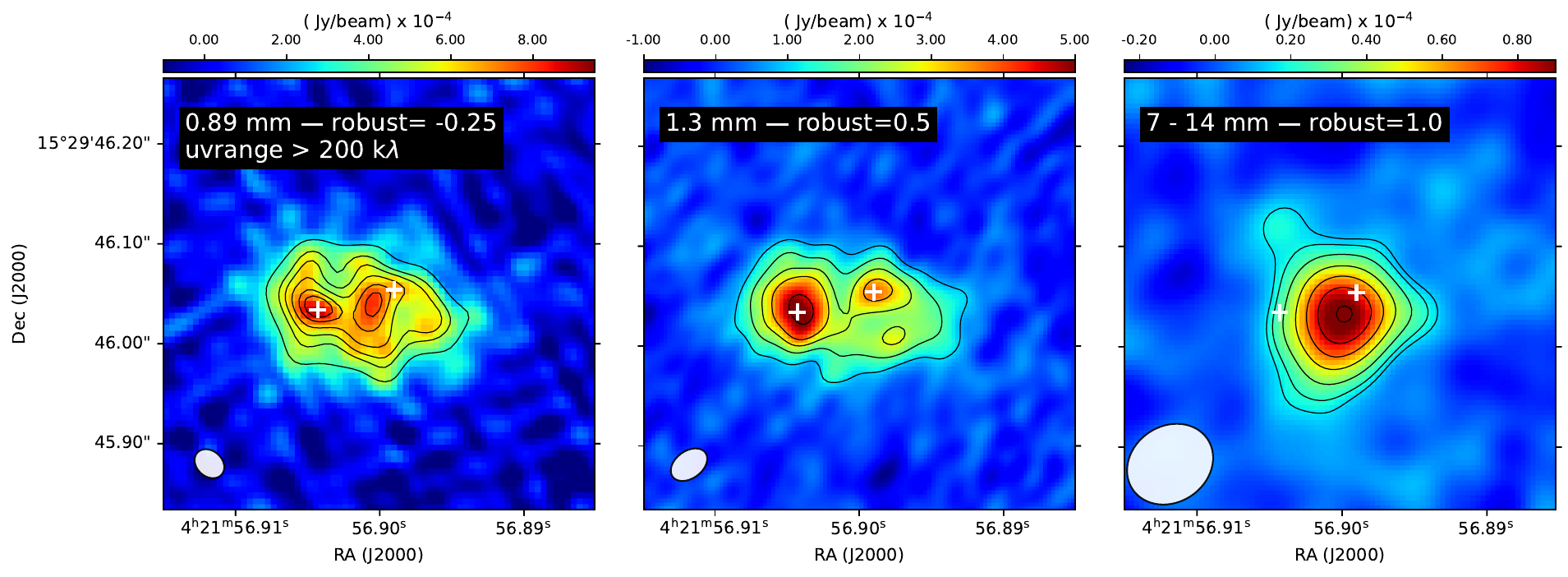}
\caption{
ALMA and VLA observations of IRAM04191.
Left: ALMA 0.89\,mm observations. The contours represent values of 
5, 7, 8, 9, 10, 10.7$\sigma$ (rms=6.6e-05 Jy/beam) and are drawn from the image smoothed with a Gaussian whose full width at half maximum corresponds to approximately one synthesized beam.
Middle: ALMA 1.3\,mm observations (robust of 0.5). The contours represent values of 4, 7, 10, 13, and 19$\sigma$, with rms= 2.4e-05\,Jy/beam. 
Right: VLA KKaQ image. The contours correspond to 3, 4, 6, 9, 12, 15, 18 times the $rms$ (5~$\mu$Jy/beam). 
The synthesized beams are displayed in the bottom-left corner of each image. The white crosses represent the positions of the two detected sources at 1.3\,mm. 
}
\label{fig1:almaima}
\end{figure*}

\begin{figure}[t!]
\includegraphics[width=0.45\textwidth]{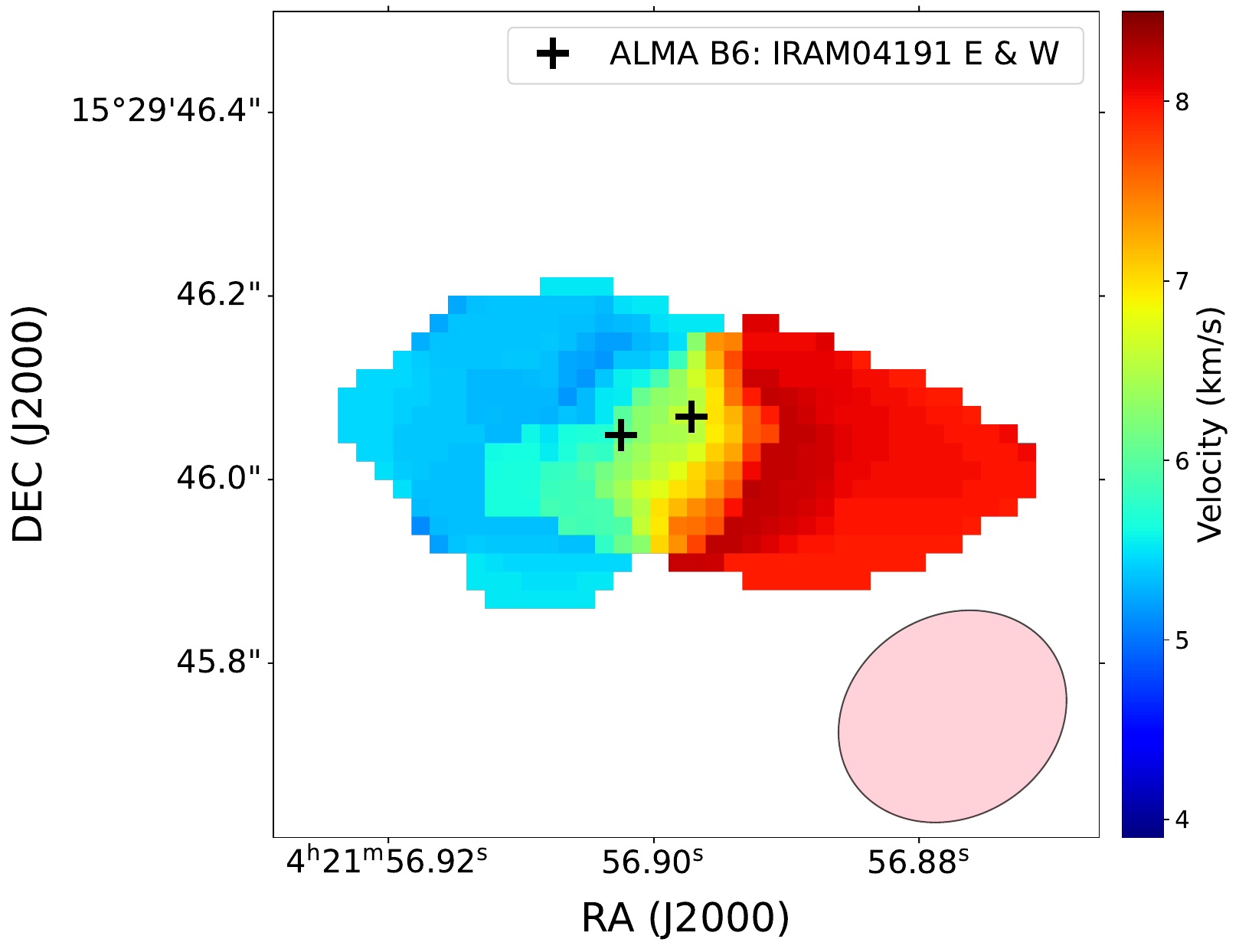}
\caption{First-order moment of the C$^{18}$O\,(2--1) line. We selected the channels that show emission closer to the central binary (see text), as they reveal the presence of rotating emission resembling a circumbinary disk. The synthesized beam is displayed at the bottom right (pink). 
The black plus signs indicate the positions of the sources detected at 1.3\,mm, as reported in Table B1.}\label{diskc18o}
\end{figure}

\begin{figure}[ht!]
\includegraphics[width=0.34\textwidth, angle=270]{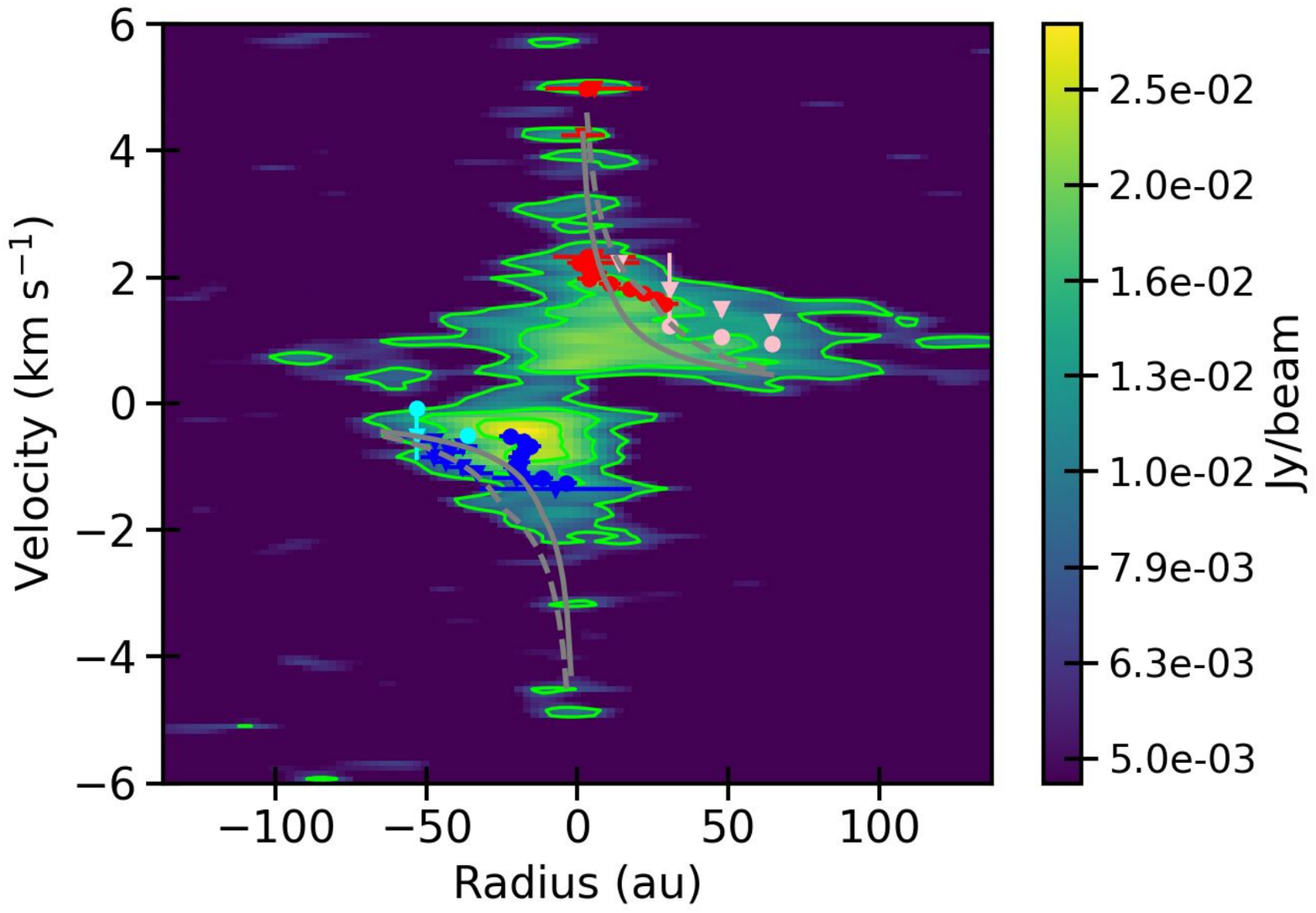}
 \caption{Position-velocity diagram built along the major axis of the \ceo\, emission. The contours represent values of  3, 6, and 9$\sigma$. 
 We include the extracted velocities from the edge (pink and cyan) and ridge (blue and red) methods within SLAM, assuming a Keplerian rotation pattern.
 }\label{fig:pvdiagram}
\end{figure}

\subsection{ALMA continuum data: Most extended configurations}\label{cont_data}

The 0.89\,mm image of IRAM04191 is displayed in Figure~\ref{fig1:almaima} (left panel). The continuum emission is extended in the east-west direction, showing two bright sources separated by $\sim$54\,mas ($\sim$8\,au at 140\,pc). The strongest emission comes from the easternmost source (IRAM04191E, hereafter). The two sources are  surrounded by extended emission. 
 
The 1.3\,mm continuum data (project 2016.1.00039.S) is displayed in the middle panel of Fig.~\ref{fig1:almaima}. It shows two emission peaks separated by $\sim$80\,mas (11\,au at 140 pc). Similar to the 0.89\,mm data, the brightest source is IRAM04191E, which is spatially resolved, while the western one (IRAM04191W) is point-like (see Table~\ref{ap:table1}). The 1.3\,mm image also reveals extended  emission connecting and surrounding the two sources. 

The positions and fluxes for the two detected sources in the two bands are included in Table~\ref{ap:table1}. We note that in the case of the 0.89\,mm data, we derived these parameters in both the self-calibrated (Fig.~\ref{fig1:almaima}) and non-self-calibrated images (see Appendix \ref{appendix_b7data} for details).

The analysis of the ALMA data suggests that IRAM\,04191 could be part of a tight binary system. This is consistent with the fact that two ALMA sources separated by $\sim$80\,mas were detected as compact sources in the two bands after a cut in the $uv$ range of 
$>$ 1100\,$k\lambda$ was applied (see Figure~\ref{fig_appendix_B7_B6_uvrange}).

We estimated the masses (gas plus dust) of the detected sources following the procedure described in Appendix~\ref{appendix_masses}. Both show very low values: between 1.8 and 0.4\,\mj\,at 1.3\,mm and below 1\,\mj\, at 0.89\,mm (see Table~\ref{ap:table1}). The mass of the whole emission (binary plus extended emission) is estimated to be 4-5\,\mj\,at 0.89\,mm and 4.1 \,\mj\,at 1.3\,mm (see details in Appendix~\ref{appendix_masses}).

\subsection{VLA observations}\label{vla_data}

The combined KKaQ-band image is displayed in the right panel of Figure~\ref{fig1:almaima}. We detected a single source at the position of IRAM04191 in the three observed bands (the three individual images are displayed in Figure\,\ref{fig:appendix_VLA}). The properties of this radio source are included in Table \ref{appendix:vla_table}. The image shows extended emission with an elongation toward the northeast in the direction of the reported outflow \citep{Andre1999}.
  Using the fluxes at 33 and 22\,GHz, we derived a spectral index of 0.24$\pm$0.23, consistent with free-free emission from a radio jet.
  
The bulk of the VLA emission seems to be associated with IRAM04191W, but the moderate angular resolution and sensitivity of the VLA data do not allow us to discern if the eastern component also contributes to the centimeter emission. The position of the VLA source is offset by $\sim$30\,mas southeast from IRAM04191W, which corresponds to the direction of the proper motion of the Taurus members \citep[e.g.][]{Galli2019_Taurus}, although it is smaller than that reported for the neighboring moving groups (e.g., L1551, L1558), considering the time span of 7 years between the ALMA and VLA data.
  However, the different beam sizes and source morphologies in the ALMA and VLA data prevent a reliable determination of the proper motion. Dedicated observations are therefore required to characterize the nature of this positional offset and the proper motion of IRAM04191.

\subsection{Molecular emission from \ceo(2-1)}
 
\citet{Maury2020} already suggested the presence of a small disk around IRAM04191, so we reanalyzed \ceo(2-1) observations to shed light on the system architecture. The moment~1 is displayed in Fig.~\ref{diskc18o}, where we selected the velocity channels that show emission close to the ALMA detections: 3.5-5.5 km/s and 7.9 – 9.4 km/s. The velocity gradient in \ceo\,is consistent with material in rotation around the ALMA binary, confirming that it may be surrounded by a circumbinary disk whose geometrical center lies closer to the faintest source (IRAM04191W). The deconvolved size of the moment~0 of the \ceo(2-1) emission is 0\farcs39$\pm$0\farcs05 $\times$ 0\farcs$17\pm$0\farcs04 (PA of 88$\pm$8$\degree$), which translates into a disk radius of $\sim$27\,au 
and an inclination\footnote{The inclination was estimated using the size ratio between major and minor disk axes following the expression {\em cos}$^{-1}$ ($\theta_{min}$/$\theta_{maj}$).} of 64$\pm$7$\degree$, which is in good agreement with the values of 50-60$\degree$derived in previous works \citep[e.g.][]{Andre1999}.

We used the moment~1 of the \ceo\,emission to generate a position-velocity ($pv$) diagram along its major axis. As shown in Fig.~\ref{fig:pvdiagram}, the diagram shows a Keplerian-like pattern.  
We fit the $pv$ diagram using the SLAM code \citep{SaiSLAM} and adopted both the edge and ridge methods. We assumed an $rms$ noise of 2.45 mJy, an S/N threshold of five, and Gaussian weighting. The ridge method provides more robust tracing of the emission, yielding a break radius of $\sim$11 au and a dynamical mass of $0.05 \pm 0.04\,M_\odot$.  

\section{Discussion}\label{discussion}

The ALMA data presented here allowed us for the first time to (i) resolve IRAM04191 into a tight Class~0 substellar binary candidate with $\sim$11\,au of projected separation surrounded by a circumbinary disk 
centered on the faintest ALMA continuum source (IRAM04191W) and to (ii) estimate a dynamical mass of $\sim$50$\pm$40\,\mj\,for the whole system. In the context of Class~0 binary systems, IRAM04191 would represent one of the tightest  binaries detected in continuum emission to date, only comparable to L1157\,MMS 
\citep[a 16\,au Class~0 binary;][]{Tobin2022_L1157}.

A two-peak morphology in the continuum emission of a young object can also be interpreted as a high-inclined disk with a large inner cavity. However, such a large cavity would be hard to reconcile with the strong molecular outflow 
reported in this source and the presence of a radio jet, which is normally associated with very young objects. Moreover, in an edge-on disk scenario, the VLA centimeter detection presented here would be expected to lie between the two ALMA peaks, and this is not the case.

Considering that IRAM04191W seems to be associated with the VLA detection and its location at the center of the circumbinary disk, it is reasonable to suggest that this source may be the primary (and hence most massive)  component of the binary, despite being the fainter emitter in the ALMA data. A comparable configuration was observed in the protostellar binary [BHB2007] 11 \citep{Alves2019}, although in 
that case both components were detected in the VLA data. Nevertheless, this hypothesis needs further confirmation through future dedicated ALMA and VLA observations.

\begin{figure}
\includegraphics[width=0.45\textwidth]
{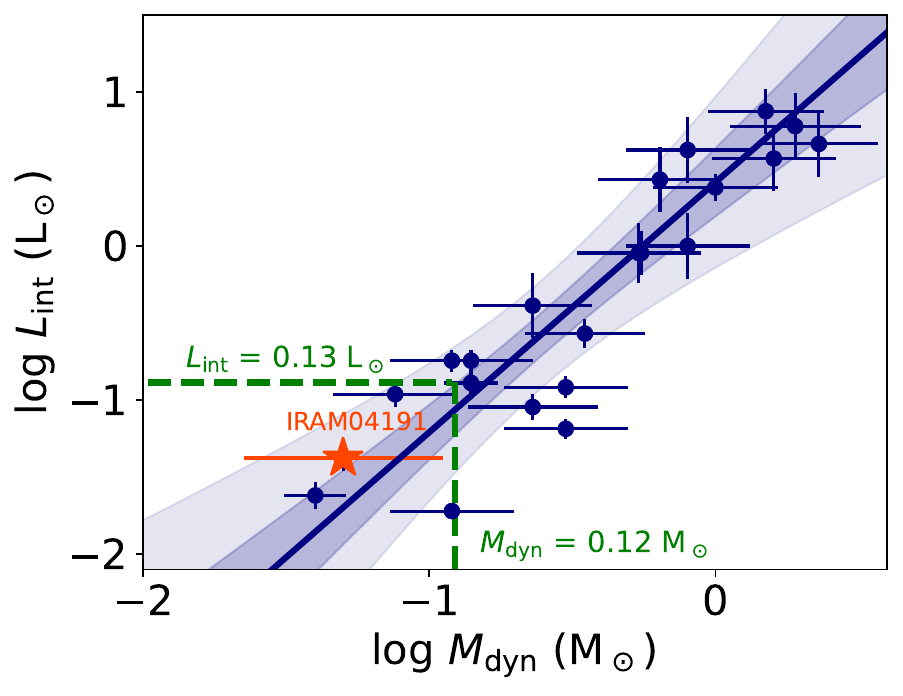}
\caption{Dynamical mass ($M_{\rm dyn}$) vs. \lint\,diagram adopted from \citet{Palau2024_review}. Blue symbols represent protostars, VeLLOs, and proto-BD candidates with \lint\, and $M_{\rm dyn}$ measurements, while the solid line represents the fit to the data. The shaded area corresponds
to two (dark) and five times (light) the uncertainty of the fit. The dashed green lines at the bottom left delimit the region consistent with proto-BDs. We represent IRAM\,04191 with $M_{\rm dyn}$ derived in this work, and \lint\,from \citet{Perez2024}. 
}
\label{fig_LintMdyn}
\end{figure}

\citet{Palau2024_review} presented a correlation between the dynamical mass ($M_{\rm dyn}$) of protostars and VeLLOs with their \lint. 
Using the dynamical mass derived in this work, we included IRAM04191 in their $M_{\rm dyn}$-$L_{\rm int}$ diagram and found it to be consistent with the reported correlation, strongly supporting its substellar nature (Fig.~\ref{fig_LintMdyn}).
If confirmed as a binary, the value of each parameter should be divided between components taking into account the binary mass ratio, which is currently unknown. 

The reported  values of $M_{\rm env}$ for IRAM04191 vary between 0.05\mo (using Herschel/SPIRE data at 250\,$\mu$m, \citealt{Kim2016}) and 0.5\,\mo\, (using 1.3\,mm data, \citealt{Andre1999}). The discrepancy between the values is mainly related to the methodology used to estimate the mass, which use a different radius (24" vs. 60", respectively) to integrate the flux. We reanalyzed the SPIRE/250\,$\mu$m data of IRAM04191 and estimated a $M_{\rm env}$ of $\sim$0.2\,\mo\, by integrating the flux within a 4$\sigma$ contour around the source and assuming $T_{\rm dust}$ and opacity as in \citet{Kim2016}. By adopting a star formation efficiency of 30\% \citep[see, e.g.,][]{Palau2024_review}, we can infer a mass reservoir of $\sim$60\,\mj\, available to be accreted by the binary over its evolution. Assuming a very conservative mass of 45\,\mj\,for each of the binary components and that each accretes half of the mass reservoir, the system might end up as a substellar pair. We note, however, that different works  have reported  episodic accretion onto the source \citep{Lee2007_vellos, Anderl2020}. This combined with the occurrence of mass-loss episodes (e.g., molecular outflow, radio jet) implies that the fate of the binary cannot be definitively established.

Early in the stellar formation process, turbulent fragmentation produced binaries with initial separations wider ($>$500\,au) than those formed by disk fragmentation, although dynamical evolution can quickly change these separations into lower values in $\sim$0.1\,Myr \citep{Offner2016}. In that respect, core fragmentation models by \citet{Bate2009, Bate2012} can produce very low mass binaries with small separations ($<$40\,au), although the modelled cloud properties do not resemble those of Taurus. An observational test that could shed light on the formation mechanism of IRAM04191 would involve studying the misalignment of the outflows and/or disks of the two components with more sensitive ALMA observations in the future \citep{Offner2016}.
Interestingly, most BD binaries observed in the field and star-forming regions are tight systems (sep $<$10-20\,au) with a mass ratio close to one \citep[e.g.][]{Duchene2013}.  The confirmation of IRAM04191 as a compact binary candidate at such an early evolutionary stage would raise the question of whether the separation distribution observed in more evolved systems is already set at very early stages of BD binary formation. 

 \begin{acknowledgements}
We are indebted to an anonymous referee that has improved the paper with a very constructive report.
This research is funded by the Spanish grant MCIN/AEI/10.13039/501100011033 PID2023-150468NB-I00. NH is grateful to ESO-Chile for hosting her through the Visitor's Program 2024, and the IRyA for hosting her through project UNAM-PAPIIT IG100223. A.P. acknowledges financial support from the UNAM-PAPIIT IN120226 grant, and the Sistema Nacional de Investigadores of SECIHTI, M\'exico.
CC-G acknowledges support from UNAM DGAPA PAPIIT grant IG101224 and from CONAHCyT Ciencia de Frontera project ID\,86372. This work benefited from the UNAM-NRAO Memorandum of Understanding in the framework of the Next Generation Very Large Array (ngVLA) Project (MOU-UNAM-NRAO-2023). N. Otten acknowledges funding from the Maynooth University Graduate Teaching Scholarship and Taighde Éireann (Research Ireland) under the RI-ESO Studentship Agreement for this work.
This paper makes use of the following ALMA data: ADS/JAO.ALMA\#2017.1.00551.S, 2016.1.00039.S, and 2016.1.01284S. ALMA is a partnership of ESO (representing
its member states), NSF (USA) and NINS (Japan), together with NRC (Canada), MOST and ASIAA (Taiwan), and KASI (Republic of Korea), in cooperation with the Republic of Chile. The Joint ALMA Observatory is operated by ESO, AUI/NRAO and NAOJ. The National Radio Astronomy Observatory and Green Bank Observatory are facilities of the U.S. National Science Foundation operated under cooperative agreement by Associated Universities, Inc.

\end{acknowledgements}

 \bibliographystyle{aa}
 \bibliography{iram04191}

\begin{appendix}
\nolinenumbers

\section{ALMA archival data reduction}\label{App:ALMA_archival_data}

The 1.3\,mm data from project 2016.1.00039.S (P.I. Dunham) were obtained on November 1, 2017, only about three weeks before our 0.89\,mm observations. The data were obtained with 49 antennas of the 12-array in single field dual-polarization mode, dedicating one spectral window of 1.875\,GHz bandwidth to observe continuum and five more spectral windows to observe spectra line emission, comprising an aggregate bandwidth of 2.262~GHz. The average precipitable water-vapor column of the observations was $\sim$0.5\,mm. The baselines ranged from 113\,m to 13.9\,km. The maximum recoverable scale was 0\farcs5. Bandpass and flux calibrations were made using QSO\,J0510+1800, while QSO\,J0431+1731 was used as a phase calibrator. The science time on-source was 31 minutes. Data were calibrated using the CASA pipeline, version 5.1.1. Continuum imaging was performed using the task \texttt{tclean} in multi-frequency synthesis mode and a multi-scale clean deconvolution algorithm with scales of 0, 0\farcs04, and 0\farcs1. A cellsize of 0\farcs004 was used, with Briggs weighting and a robust parameter = 0.5, providing a synthesized beam of 0\farcs04$\times$0\farcs03 and PA of -36\degree. We tried the self-calibration of the data, but it did not work due to the faintness of the source. The uvrange of the data spans from 53 to 9200$k\lambda$.

We have also analyzed data from project 2016.1.01284.S (P.I. Maury), which were obtained in two executions on October 15, 2016, and July 21, 2017.  Observations were taken with 41 and 46 antennas of the 12-array respectively, in single field dual-polarization mode. Three spectral windows of 1.875 GHz bandwidth each were set up to observe continuum, and four more spectral windows of 58.594 MHz bandwidth were selected to observe spectra line emission, comprising an aggregate bandwidth of 5.859 GHz. The average precipitable water-vapor column of the observations was $\sim$0.64\,mm. The baselines ranged from 16.7\,m to 3.7\,km. The maximum recoverable scale was 1\farcs9. Bandpass calibration were made using QSO\,J0425+1755 and J0237+2848.  QSO\,J0425+1755 and  J0423-0120  were used for flux calibration, and QSO\,J0433+0521 and J0449+1121 were used as phase calibrators respectively. The total time on the science source was 1.1~hr. Data were calibrated using version 4.7.2 of the CASA pipeline.  
For this paper, we have only analyzed the spectral window that covered the emission at C$^{18}$O (2-1), with 61.035 kHz spectral resolution, equivalent to 0.08\,km/s velocity resolution.
We note that this dataset was already discussed in \citet{Maury2020}, who already suggested the presence of a small disk around the source. We have reprocessed the data to isolate the disk emission in order to (i) study its location with respect to the binary components, and (ii) to estimate a dynamical mass for the whole system.
C$^{18}$O(2-1) imaging was performed using the task \texttt{tclean} using hogbom cleaning deconvolution algorithm. A cell size of 0\farcs02 was used, with Briggs weighting and a robust parameter = 1.5, providing a synthesized beam of 0$\farcs$3$\times$0\farcs2, and PA of -57.8$\degree$. CASA version 6.5.6 was used for imaging purposes. 

\section{ALMA Band 7 data}\label{appendix_b7data}

\begin{figure*}
\centering
\includegraphics[width=0.84\textwidth]{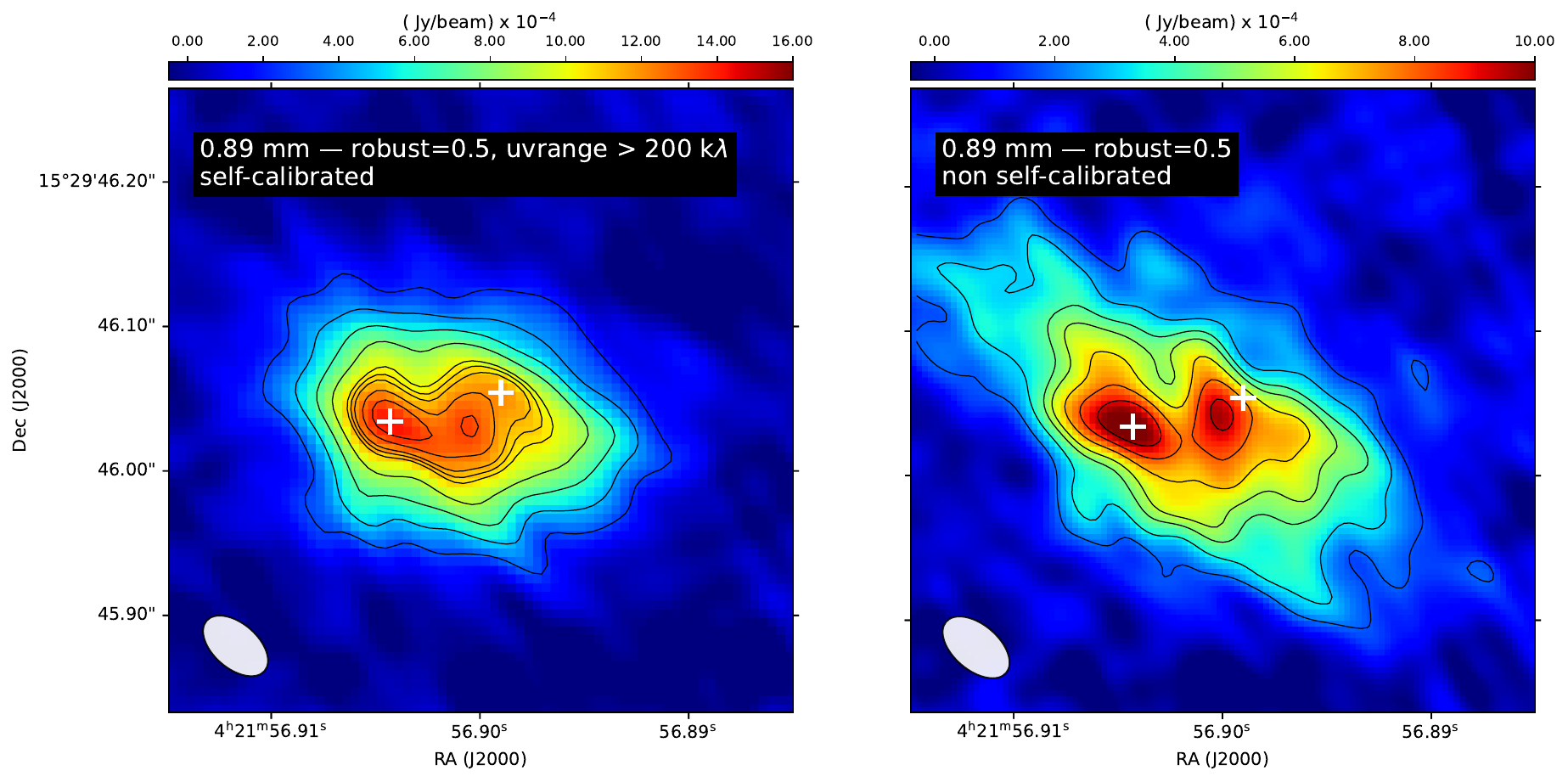}
\caption{ALMA 0.89\,mm observations of IRAM04191 with a robust of 0.5. Left: Self-calibrated image. 
The contours represent 5, 9,13, 17, 21, 22, 23, 25, 27 times the $rms$ (4.9e-5 Jy/beam). 
Right: Non-self-calibrated image. The contours represent 
4, 7, 10, 12, 15, 18, 21 times the $rms$ (4.5e-5 Jy/beam). 
The white crosses represent the position of the two sources detected in the 1.3\,mm data.
}\label{fig_appendix_self}
\end{figure*}

To further investigate the presence of two compact sources in the ALMA data, we compared images generated using different robust
and a cut in uvrange $>$200 $k\lambda$.
We have explored robust of 0.5, 0.3, -0.25, -0.5 and -1, and we
display the images with -0.25 and 0.5 in Figs. \ref{fig1:almaima} and
\ref{fig_appendix_self}, respectively. The left panel of Fig.~\ref{fig_appendix_self} shows the 0.89\,mm image with robust 0.5, in which we detect a central source resolved into two bright peaks separated by 0\farcs04,  and an additional  extended emission toward the northwest, in the direction where we detect the secondary component in the 1.3\,mm data. As the robust is decreased, giving more weight to the longest baselines, the two central peaks set apart, and show a separation of $\sim$0\farcs054 for a robust of -0.25, as shown in Section \ref{alma:b7}. 
Although this particular robust allows us to resolve the binary, it has the disadvantage of increasing the image noise ($rms$ 6.6e-5 Jy/beam) in comparison with the robust of 0.5 (4.9e-5 Jy/beam). 

For comparison, the panel on the right of Fig.~\ref{fig_appendix_self} shows the non-self-calibrated 0.89\,mm image obtained using a robust of 0.5. For this data, the calibration was done with CASA version 5.1.1. To  produce the continuum image shown in the figure, we used a more recent version of CASA (6.5.6).  The task \texttt{tclean} was used in multifrequency synthesis mode, with multiscale clean deconvolution algorithm comprising scales of 0, 0\farcs02, and 0\farcs1. We used a cellsize of 0\farcs004, with Briggs weighting and robust parameter of 0.5, providing the best trade-off between spatial resolution and sensitivity. The resulting synthesized beam is 0\farcs05$\times$0\farcs03, and position angle of 40.4$\degree$. This image allows us to recover the two components of the binary system more easily because the extended emission is decorrelated and artificially spread along the northeast-southwest direction, revealing more clearly the central emission. The resulting image shows an $rms$ of 4.5e-5 Jy/beam, better than the value obtained in the self-calibrated image with robust -0.25. Hence,  we have also used this image to derive the physical parameters of the binary components included in Table \ref{ap:table1}.

\begin{figure*}
\centering
\includegraphics[width=0.84\textwidth]{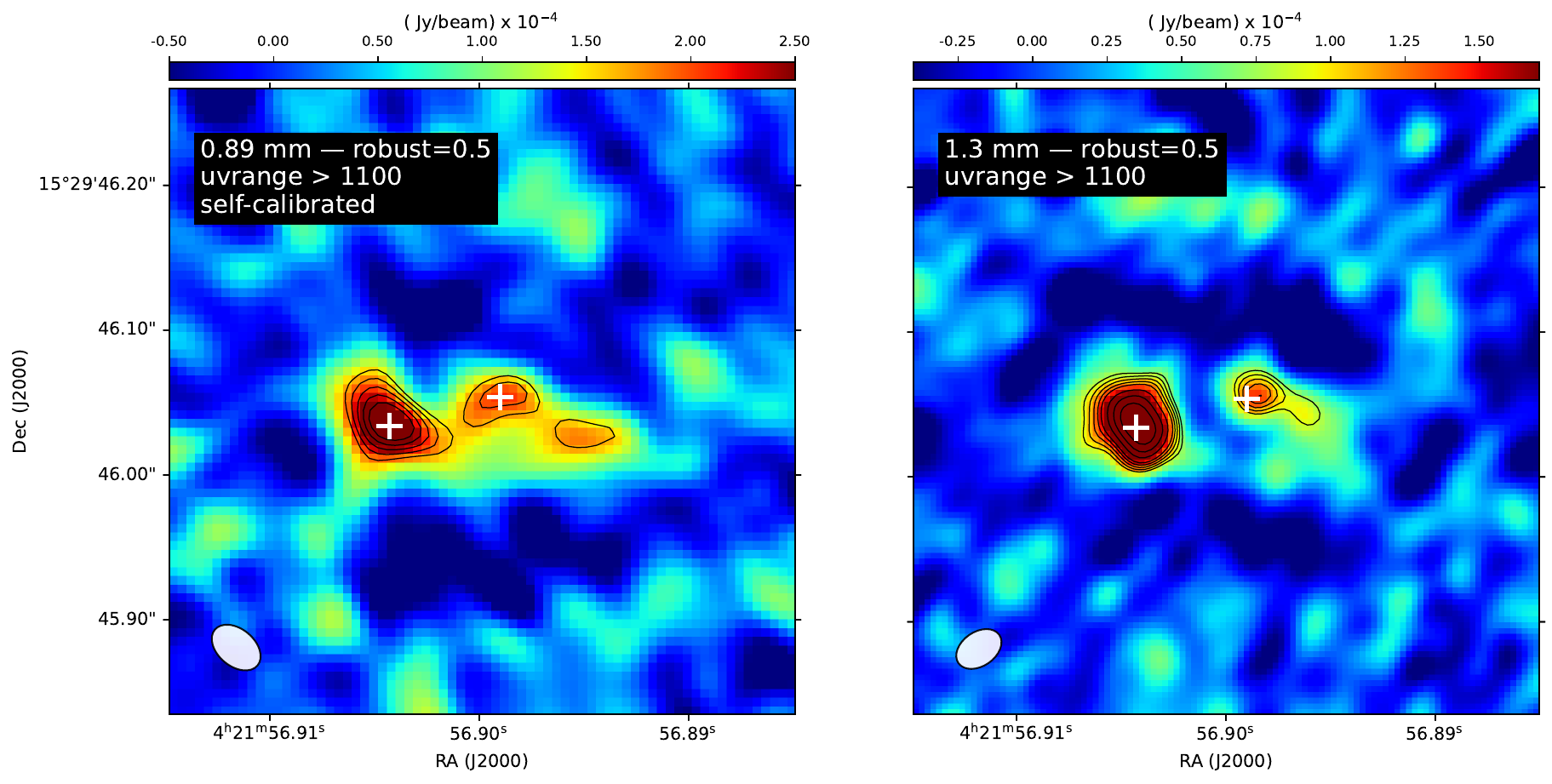}
\caption{Observations of IRAM04191 with $uv$ range $>$1100\,k$\lambda$. The images are smoothed with a Gaussian whose full width at half maximum is approximately equal to the synthesized beam. The contours in the two images represent 3, 3.5, 4, 4.5, 5, 6, 7, 8 times the $rms$. Left panel: ALMA 0.89\,mm data ($rms$ 5.4e-5 Jy/beam). Right panel: ALMA 1.3\,mm data ($rms$ = 2.8e-5\,Jy/beam). The white crosses represent the position of the two sources detected in the ALMA 1.3\,mm data with full $uv$ range displayed in Fig. \ref{fig1:almaima}.
}\label{fig_appendix_B7_B6_uvrange}
\end{figure*}

 Due to the fact that a negative robust enhances artificial clumpy emission, we have also generated a 0.89\,mm self-calibrated image with robust of 0.5 and a cut in $uv$ range of $>$1100 $k\lambda$, giving more weight to the most extended baselines. The result is displayed in Fig.~\ref{fig_appendix_B7_B6_uvrange} (left panel). The image with the $uv$ cut shows a strong source to the east plus an additional well-separated source to the west at a separation of $\sim$80\,mas. The two bright emissions  are connected by a stream of material. For comparison, we have also applied a cut in $uv$ range of $>$1100 $k\lambda$ to the 1.3\,mm image (\ref{fig_appendix_B7_B6_uvrange}, right panel), recovering the reported binary with the detection of two compact sources at a separation of $\sim$80\,mas.

\section{ALMA parameters of the sources detected at 0.89\,mm and 1.3\,mm continuum images}\label{App:ALMA_detections}

Table~\ref{ap:table1} includes the coordinates,  peak intensities, fluxes, and deconvolved sizes of the two sources detected in the 0.89\,mm and 1.3\,mm ALMA images.
Note that, for the reasons explained above, the 0.89\,mm measurements have been performed both in the self-calibrated image shown in Figure \ref{fig1:almaima}, and in the non self-calibrated image displayed in the right panel of Figure \ref{fig_appendix_self}.
The ALMA flux calibration accuracy is $\sim$10$\%$ in the two bands (see, e.g., 10.5281/zenodo.4511522). 
Positions and peak intensities of the two sources have been derived using \texttt{imstat}. A double Gaussian has been fit to the most compact emission of the data to estimate the flux density and the deconvolved size of the two binary components in the two bands.

\begin{table*}
\caption{Properties of the  ALMA detected emissions toward IRAM04191.}
\label{ap:table1}
\begin{tabular}{lccccccc}\hline
Source & RA(J2000) & DEC(J2000) & Peak$^{1}$ & Flux$^2$  & Deconvolved$^{3}$ & Mass \\
   & & &  Intensity& Density & size & \\
       & (hh:mm:ss) & (dd:mm:ss)  & (mJy/beam) & (mJy) & (mas, $\degree$) & (M$_{\rm Jup}$) \\ 
\hline \\ [-0.7em]

\multicolumn{7}{c}{Band 7 (2017.1.00551.S) -- self-calibrated image} \\ 
\hline \\[-0.7em]
IRAM04191E & 04:21:56.9044 & 15.29.46.035 & 0.91$\pm$0.07 & 1.63$\pm$0.19 & 37.8$\pm$0.2$\times$15.7$\pm$0.4, PA=49$\pm$7 & 0.45 \\ 
IRAM04191W & 04:21:56.9007 & 15.29.46.041 & 0.81$\pm$0.07  & 1.16$\pm$0.15 & point-like &  0.32\\
\hline \\ [-0.7em]
\multicolumn{7}{c}{Band 7 (2017.1.00551.S) -- non self-calibrated image} \\ 
\hline \\ [-0.7em]
 IRAM04191E  & 04:21:56.9046 & 15:29:46.034 & 1.07$\pm$0.04 & 1.98$\pm$0.10  & 58.43$\pm$0.08 $\times$ 24.5$\pm$0.2, PA=43$\pm$3 &  0.54\\
  IRAM04191W & 04:21:56.9001 & 15:29:46.042  & 0.97$\pm$0.04 & 1.28$\pm$0.08 & 37.4$\pm$0.8 $\times$ 8.2$\pm$6.0, PA=14$\pm$5 & 0.35 \\  
\hline \\ [-0.7em]
   \multicolumn{7}{c}{Band 6 (2016.1.00039.S)} \\ 
   \hline \\ [-0.7em]
    IRAM04191E & 04:21:56.9043 & 15:29:46.034  & 0.52$\pm$0.02 & 1.89$\pm$0.06 & 64.5$\pm$0.2 $\times$ 41.4$\pm$0.3, PA=54$\pm$10  & 1.75\\
    IRAM04191W & 04:21:56.8990 & 15:29:46.054 &  0.38$\pm$0.02 & 0.48$\pm$0.04 & point-like  & 0.45\\
\hline
\end{tabular}
Notes to table: 
$^{1}$ The reported error is the $rms$ of the map; 
$^{2}$ Flux density inferred from a double Gaussian fitting to the most compact emission of the data; 
$^{3}$ Estimated from a double Gaussian fitting to the data. We provide the size and the PA.
\end{table*}

\section{Estimation of masses from ALMA continuum data}\label{appendix_masses}

  We have estimated the mass of the ALMA sources 
 assuming optically thin and isothermal dust emission, and 
 following the expression:
 
\begin{equation}\label{app:equation_mass}
M_{\rm dust} = \frac{F_{\nu} d^2}{ k_{\nu} B_{\nu}(T_{\rm d})},
\end{equation}

where $F_{\nu}$ is the flux density at the given frequency, $d$ is the distance to the source, $k_{\nu}$ is the dust mass opacity, and $B_{\nu}(T_{d})$ is the Planck function at the dust temperature $T_d$. We have assumed a dust temperature of 12.5\,$K$ \citep{Andre1999}, opacities of 1.68 and 0.89 cm$^2$/g for 890\,$\mu$m and 1.3\,mm (for thin ice mantles and density of 10$^{-6}$ cm$^3$), respectively \citep{Ossenkopf1994}, and a gas-to-dust mass ratio of 100. The masses are included in the last column of Table \ref{ap:table1}.
We have also estimated the mass of the whole integrated emission (binary plus extended emission) in both bands. 
 To do so, we have first integrated the whole emission within a 3$\sigma$ contour in the images, resulting in density fluxes of 14.3$\pm$0.4 (self-calibrated, Fig.~\ref{fig1:almaima}) and 17.4$\pm$0.2\,mJy (non self-calibrated, right panel of Fig. \ref{fig_appendix_self}) in the 0.89\,mm images,  and 4.4$\pm$0.1\,mJy at 1.3\,mm. Using equation \ref{app:equation_mass}, we derive masses of $\sim$4\,\mj\,and 
 $\sim$5\,\mj\,for the self- and non-self-calibrated images at 0.89\,mm, and  $\sim$4\,\mj\,at 1.3\,mm.

\section{VLA observations of IRAM04191}\label{App:VLA}

\begin{figure*}
\includegraphics[width=0.97\textwidth]{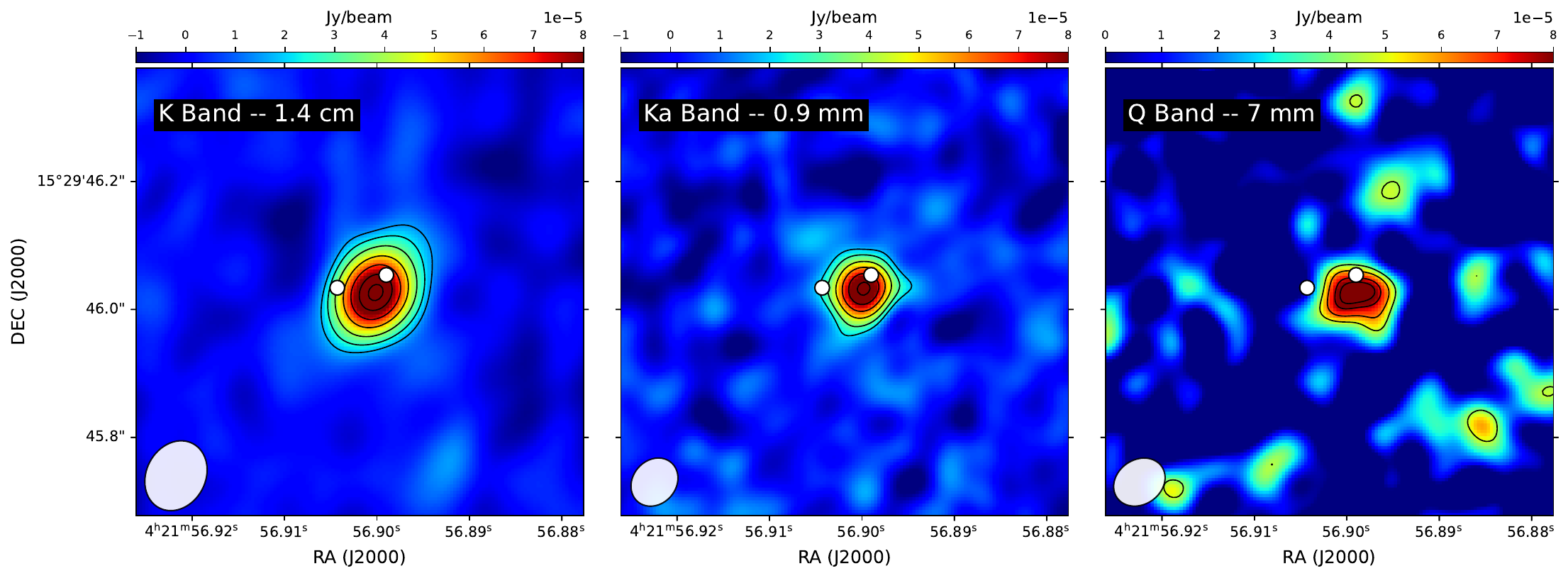}
\caption{VLA images of IRAM04191 in the three observed bands: 1.4\,cm (left),  9\,mm (middle), and 7\,mm (right). 
The first panel shows contours at 
4, 6, 9, 13, 17, 21, 24 $\times$ the $rms$ (4 $\mu$Jy/beam) and the second panel 4, 6, 9, 12, 15, 18 $\times$ $rms$ (5\,$\mu$Jy/beam). The 7\,mm image displays contours representing 3, 4, 5,  and 5.5 $\times$ $rms$ (15\,$\mu$Jy/beam). Natural weighting is used in all three images; the 7 mm image includes a 2000 $k\lambda$ taper.
}\label{fig:appendix_VLA}
\end{figure*}
\begin{table*}
\caption{Properties of the VLA detected source.}\label{appendix:vla_table}
\begin{tabular}{lcclccccc}
\hline
Band & $\nu$ & $\lambda$ & RA(J2000)  & DEC(J2000) & Peak  & Flux& Deconvolved & Beam \\ 
     & & & & & Intensity & Density & Size$^1$ & Size$^1$ \\ 

& (GHz) & (cm) & (hh:mm:ss) &    (dd:mm:ss)    & ($\mu$Jy/beam) & ($\mu$Jy) & (mas, $\degree$) & (mas, $\degree$) \\ 
     \hline
K    & 22& 1.40 & 04:21:56.900 & 15:29:46.026 & 98$\pm$3 & 118$\pm$6 & 60$\times$40$\pm$10, PA=144$\pm$30 & 113$\times$92, PA=-27.9 \\ 
Ka   & 33& 0.90 & 04:21:56.900 & 15:29:46.030 & 93$\pm$5 & 130$\pm$10 & 44$\times$42$\pm$20, PA=144$\pm$30 & 80$\times$68, PA=-40.7 \\
Q    & 44& 0.68 & 04:21:56.900 & 15:29:46.020 & 93$\pm$7 & 180$\pm$20 & 90$\times$50$\pm$20, PA=70$\pm$20 & 85$\times$70, PA=-55.0\\
\hline
\end{tabular}

Notes: For K and Ka, the measurements were obtained on the images with natural weighting, while for Q they were measured on the images with natural weighting and a taper of 2000\,$k\lambda$. $^1$ We provide the size in milliarcseconds and the PA in degrees, with their associated uncertainties. $^2$ We provide the beam size and its PA in each band.
\end{table*}

We have detected a single source in the VLA observations presented in Section~\ref{vla_data}. Figure~\ref{fig:appendix_VLA} shows the individual images in the K, Ka, and Q bands, while 
Table \ref{appendix:vla_table} summarizes the properties of the detected source in the three bands. Note that the flux calibration uncertainties in K, Ka and Q-bands are of 10-12\%\footnote{https://science.nrao.edu/facilities/vla/docs/manuals/oss/performance/fdscale}.

\end{appendix}
\end{document}